\documentstyle[aps,psfig]{revtex}

\newcommand{\be}[1]{\begin{equation} \label{(#1)}}
\newcommand{\ee}{\end{equation}}
\newcommand{\ba}[1]{\begin{eqnarray} \label{(#1)}}
\newcommand{\ea}{\end{eqnarray}}

\begin{document}

\title{Electromagnetic and weak hyperon properties in the Skyrme model
\footnote{Cont. to the Proc. of the "School on Electromagnetic probes and the structure of hadrons and nuclei".
Erice, Italy. September 17- 24, 1999 to be published in Progress in Particle and Nuclear Physics.}}

\author{Norberto N. Scoccola$^{a,b}$}

\address{
$^a$ Physics Dept., Comisi\'on Nac. de Energ{\'{\i}}a At\'omica,
Libertador 8250 (1429) Bs.As., Argentina.\\
$^b$ Universidad Favaloro, Sol{\'{\i}}s 453 (1078) Bs. As., Argentina.}

\maketitle

\begin{abstract}
We report on the result of some investigations concerning the radiative decays of decuplet baryons and the
non-leptonic weak decays of the octet baryons in the context of topological chiral soliton models. Our results are
compared with those of alternative baryon models. For the radiative decays we find that the predictions are similar
to those of quark models. In the case of the non-leptonic weak decays, we find that although the predicted $S$-wave
amplitudes are in rather good agreement with the observed values, the model is not able to reproduce the empirical
$P$-wave amplitudes. Thus, in contrast to previous expectations, the Skyrme model does not seem to provide a
solution to the long-standing '$S$-wave/$P$-wave puzzle'.
\end{abstract}

\section{Introduction}

During the last fifteen years it has become clear that the applicability
of the Skyrme's topological soliton model
for baryon structure\cite{Sky61} goes beyond all the original expectations. Indeed, this model
provides a very useful
framework to investigate not only the structure of non-strange baryons and
their interactions\cite{ZB86}, but of strange hyperons
as well. In this type of models, baryons emerge as soliton configurations of
the pseudoscalar mesons. The corresponding non-linear chiral
action describes the main features of QCD at low energies. When the model
is extended to the strange sector, and in order to account for
the non-negligible strange quark mass, appropriate chiral symmetry
breaking terms have to be included.
One can treat the resulting effective action by starting
from a flavor symmetric formulation wherein non--vanishing kaon fields arise from a
rigid rotation of the classical pion field. The associated collective
coordinates, which parametrize these large amplitude fluctuations
off the soliton, are canonically quantized to generate states which
possess the quantum numbers of physical hyperons. It turns out that
the resulting collective Hamiltonian can be diagonalized exactly even
in the presence of flavor symmetry breaking \cite{YA88}.
This approach leads to a good description
of the hyperon spectra, magnetic moments, etc (for a review see
Ref.\cite{Wei96}). Consequently, it provides an alternative and interesting
scheme to investigate other various hyperon properties.
In the present contribution we will concentrate
on the application of the model to the study of the electromagnetic
decays of the decuplet baryons and of the non-leptonic weak decays
of the octet hyperons.

The study of the decuplet baryon radiative decays
is of great interest at the present time due to the upcoming
experiments at JLAB\cite{JLAB}. In fact, although recently the reaction
$\Delta\to N\gamma$ has carefully been analyzed at MAMI, the decay parameters
are still unknown for those $J=\frac{3}{2}$ to $J=\frac{1}{2}$ transitions which
involve strange baryons. The experiments at JLAB are expected to
provide some data on these
radiative decays soon and thus give more insight in the
pattern of flavor symmetry breaking. Theoretical studies of these
decays have been performed in a number of quark-based models
like i.e. the non--relativistic quark model \cite{Da83,Le93},
a quenched lattice calculation \cite{Le93}, etc. Here, we report
the corresponding results in the Skyrme model.
On the other hand, the non-leptonic weak
decays of hyperons represent a long-standing problem
in hadron physics. For example, although quark models with QCD enhancement
factors have been quite successful in predicting $S$-wave decay
amplitudes (see Ref.\cite{DGH86} and references therein),
they have serious difficulties in reproducing
the empirical $P$-wave amplitudes. Indeed, this
is a problem (so-called "$S$-wave/$P$-wave puzzle") which seems to be
common to other approaches as e.g. heavy baryon chiral
perturbation theory\cite{Jen92}. Within the context of the
topological soliton models a possible solution to this problem has
been suggested rather long ago\cite{DGL85}. It was shown that in the Skyrme
model, in addition to the standard pole terms, the $P$-wave
amplitudes receive extra contributions from contact terms.
Unfortunately, at the time this suggestion was made the model was
hampered by several serious problems, like e.g. very poor
predictions for the hyperon spectrum\cite{Che85}, far too small
results for the $S$-wave non-leptonic decay
amplitudes\cite{DGL85,PT85}, etc. Consequently, it was hard to draw
definite conclusions about the real relevance of such contact
terms. As mentioned above the introduction of a more refined version
of the model has led to the
solution of such problems. In fact, it has been recently
shown\cite{Sco98} that the correct $S$-wave absolute values can be
naturally obtained within such scheme.  Thus, we are now in
position to verify whether chiral soliton models can provide a
unified and consistent description of the hyperon non-leptonic
decays. The strategy is to use an effective chiral lagrangian
to describe the weak interactions. Assuming that the corresponding
low energy constants are fixed to describe the known $K$ meson
decay amplitudes, the hyperon weak matrix elements are then evaluated using
the topological soliton model wave functions.

The rest of this contribution is organized as follows. In Sec. 2
we give a brief description of the collective coordinate approach
to the $SU(3)$ Skyrme model. In Sec. 3 we present our predictions
concerning the radiative decays of the decuplet baryons within
this approach. In Sec. 4 we describe how to evaluate the
non-leptonic hyperon decays and give our results for the
$S$-wave and $P$-wave decay amplitudes. Finally in Sec. 5 some
conclusions are given.

\section{The model}

As mentioned in the Introduction, in the Skyrme model baryons appear as
topological excitations of a chiral effective action which depends only on meson
fields. In the present work, the following effective action is used
\begin{equation}
\Gamma = \Gamma_{SK} + \Gamma_{WZ} + \Gamma_{SB} \ ,
\label{action}
\end{equation}
where $\Gamma_{SK}$ is the Skyrme action
\begin{equation}
\Gamma_{SK} =
\int d^4 x \left\{ {f^2_\pi \over 4}
\mbox{Tr}\left[ \partial_\mu U (\partial^\mu U)^\dagger \right]
+
 {1\over{32 \epsilon^2}}
 \mbox{Tr}\left[ [U^\dagger \partial_\mu U , U^\dagger \partial_\nu U]^2\right] \right\} \, .
\end{equation}
Here, $\epsilon$
is the dimensionless Skyrme parameter. Furthermore the chiral field
$U$ is the non--linear realization of the pseudoscalar octet.
$\Gamma_{WZ}$ is the Wess-Zumino action
\begin{eqnarray}
\Gamma_{WZ} &=& - {i N_c \over{240 \pi^2}}
\int d^5x \  \epsilon^{\mu\nu\rho\sigma\tau} \
\mbox{Tr}[ L_\mu L_\nu L_\rho L_\sigma L_\tau] \ ,
\end{eqnarray}
where $L_\mu = U^\dagger \partial_\mu U$ and $N_c=3$ is the number of colors.
Finally, $\Gamma_{SB}$ represents the symmetry breaking terms
\begin{eqnarray}
\Gamma_{SB} & = &\int d^4x \left\{
 { f_\pi^2 m_\pi^2 + 2 f_K^2 m_K^2 \over{12} }
 \mbox{Tr} \left[ U + U^\dagger - 2 \right]
+ \sqrt{3}  { f_\pi^2 m_\pi^2 - f_K^2 m_K^2 \over{6} }
\mbox{Tr} \left[ \lambda_8 \left( U + U^\dagger \right) \right] \right.
\nonumber \\
& & \qquad
 \left.
+ { f_K^2 - f_\pi^2\over{12} }
\mbox{Tr} \left[ (1- \sqrt{3} \lambda_8)
\left(
U (\partial_\mu U)^\dagger \partial^\mu U +
U^\dagger \partial_\mu U (\partial^\mu U)^\dagger \right)
\right] \right\} \, .
\label{sb}
\end{eqnarray}
\noindent
Here, $f_K$ is the kaon decay constant while $m_\pi$ and
$m_K$ are the pion and kaon masses, respectively. In all our
numerical calculations below we will take $\epsilon = 4$ and
the rest of the parameters appearing in $\Gamma$ to their
empirical values.

In the soliton picture we are using the strong interaction
properties of the low--lying $\frac{1}{2}^+$ and $\frac{3}{2}^+$
baryons are computed following the standard $SU(3)$ collective
coordinate approach to the Skyrme model. We introduce the
ansatz
\begin{equation}
U_0({\bf r}, t) = A(t) \ \left(
\begin{array}{cc}
\cos F(r) + i \mbox{\boldmath $\tau$} \cdot
{\hat{\mbox{\boldmath $r$}}} \ \sin F(r) & 0 \\
0  & 1
\end{array}
\right)
\ A^\dagger(t) \
\label{ansatz}
\end{equation}
for the chiral field.
Here, $F(r)$ is the chiral angle which parameterizes
the soliton. The collective rotation matrix $A(t)$ is $SU(3)$ valued.
Substituting the configuration Eq.(\ref{ansatz}) into $\Gamma$
yields (upon canonical quantization of $A$) the collective hamiltonian
\begin{equation}
H = M_{sol} +\left(\frac{1}{2\Theta_\pi}-
\frac{1}{2\Theta_K}\right)  {\mbox{\boldmath $J$}}^2
+ \frac{1}{2\Theta_K} C_2\left[ SU(3)\right]-\frac{3}{8\Theta_K}
+ \Phi \ \left( 1 - D_{88} \right) \ .
\label{slowham}
\end{equation}
In Eq (\ref{slowham}), $\Theta_\pi$ and $\Theta_K$ are the moments
of inertia in the pionic and kaonic directions, respectively, and
$\Phi$ the symmetry breaking strength (see Ref.\cite{Wei96} for details).
$\mbox{\boldmath $J$}$ denotes the spin operator
while $C_2\left[ SU(3)\right]$ refers to
the quadratic Casimir operator of $SU(3)$.
The eigenfunctions and eigenvalues of the collective Hamiltonian are identified
as the baryon wavefunctions $\Psi_B(A)=\langle B |A\rangle$ and masses $m_B$.
Due to the symmetry breaking terms in $\Gamma_{SB}$ this Hamiltonian
is obviously not $SU(3)$ symmetric. As shown by Yabu and Ando \cite{YA88}
it can, however, be diagonalized exactly. This diagonalization
essentially amounts to admixtures of states from higher dimensional
$SU(3)$ representations into the octet ($J=\frac{1}{2}$) and decuplet
($J=\frac{3}{2}$) states. This procedure has proven to be quite successful
in describing the hyperon spectrum and static properties \cite{Wei96}.

\section{Radiative decay widths}

To study the hyperon radiative decays one has to
introduce the electromagnetic field $A_\mu$ in the effective
action described in the previous section. This is accomplished
by replacing the usual derivatives $\partial_\mu$ appearing in
$\Gamma_{SK}$ and $\Gamma_{SB}$ by covariant derivatives.
These covariant derivatives are defined as
\begin{equation}
D_\mu U=\partial_\mu U + ie\ A_\mu \ [{\cal Q}, U ] \ ,
\end{equation}
where ${\cal Q}$ is the electric charge matrix
${\cal Q} = (\lambda_3 + \lambda_8/\sqrt{3} )/2$.
In the case of Wess-Zumino action $\Gamma_{WZ}$, the gauging procedure is
slightly more complicated since extra terms have to be added in order to
ensure gauge invariance. Details of this procedure as well
as the explicit form of the gauged Wess-Zumino action
can be found in e.g. Ref.\cite{Wit83}. The resulting gauged effective
action can be cast in the following form
\begin{equation}
\Gamma^{gauged} = \Gamma + e \int d^4x \ A_\mu \ J^\mu_{em} +
e^2 \int d^4x \ A_\mu \ G^{\mu\nu} \ A_\nu \ ,
\label{gauged}
\end{equation}
where $J^{em}_\mu$ is the hadron e.m. current and $G^{\mu\nu}$ is
a tensor which is responsible, for example, of the seagull contributions
to the hadron polarizabilities. For the calculation of the radiative
decays only the second term in Eq.(\ref{gauged}) is relevant.

For the e.m. decays of the decuplet hyperons we are interested in
both $M1$ and $E2$ transitions are allowed. The corresponding partial
widths can be expressed in terms of $J^{em}_\mu$
as
\begin{eqnarray}
\Gamma_{E2}&=&\frac{675}{8}e^2 q
\left|\langle \Psi_{J=\frac{1}{2}}
|\int d^3r \ j_2(qr)\left(\frac{z^2}{r^2}-\frac{1}{3}\right)J^{em}_0|
\Psi_{J=\frac{3}{2}}\rangle\right|^2
\label{GE2}  \\
\Gamma_{M1}&=&18e^2 q
\left|\langle \Psi_{J=\frac{1}{2}} |\frac{1}{2}\int d^3r \ j_1(qr)
\ \epsilon_{3ij}\ {\hat r}_i \ J^{em}_j |
\Psi_{J=\frac{3}{2}}\rangle\right|^2 \ ,
\label{GM1}
\end{eqnarray}
where $q$ is the photon momentum and $j_l(qr)$ are the spherical
Bessel functions.

In our model, the general form of $J_\mu^{em}$ can be directly
obtained from the gauged effective action.
Using the ansatz Eq.(\ref{ansatz}) one can then get the explicit
expression of the integrands in Eqs.(\ref{GE2},\ref{GM1}). These
expressions can be found in Ref.\cite{HRSW97}. It should be noticed
that in deriving Eq.(\ref{GE2}) some approximations according to
the Siegert's theorem have been made.
The impact of such approximations on the E2$-\Delta N$
transition has been recently studied in the two flavor reduction of
the model\cite{WH97}. It turns out that the kinematical corrections
coming from the low-momentum expansion of the Bessel functions
are of the order of 5 \%.
In addition to this, pion fluctuations off the rotating soliton
have to be included to consistently satisfy the continuity equation
$\partial_\mu J^\mu=0$ at subleading order in $1/N_C$. These induced
fields account for shortcomings in the
collective quantization and give corrections of
$\sim 25$\% for the $\Delta N$ case. Unfortunately such an inclusion
of induced fields seems to be unfeasible in the three flavor model with
symmetry breaking included. In any case, since the decuplet decays
are almost completely M1--dominated, these uncertainties do not affect
the total decay widths in any significant way.

Results for the total decay widths are given in Table 1.
There we list
the corresponding widths normalized to the $\Delta N$ value. As in
the case of magnetic moments, the calculated $\Delta N$ transition
amplitude turns to be roughly $30 \%$ smaller than the empirical
value. However, it has been recently shown that in the $SU(2)$ model
the inclusion of next-to-leading order quantum corrections solves not only
the nucleon mass problem, but also that of
the magnetic moments\cite{MW97}. Similar improvements are expected
in the present model.
From Table 1 we observe that the predictions of the present
model  are very similar to those of
quark-based models. In particular, in both types of models the
$U-$spin selection rules that predict small
$\Sigma^{*-} \rightarrow\gamma\Sigma^-$ and
$\Xi^{*-}    \rightarrow\gamma\Xi^-$
are rather well satisfied.

\begin{table}[h]

\begin{center}

\begin{tabular}{lccc}
Transition & This model & QM & LATT.  \\ \hline $\Sigma^{*0} \rightarrow\gamma\Lambda  $ & 0.653 & --    & 0.703 \\
$\Sigma^{*-} \rightarrow\gamma\Sigma^- $ & 0.007 & 0.007 & 0.006 \\ $\Sigma^{*0} \rightarrow\gamma\Sigma^0 $ &
0.035 & 0.040 & 0.055 \\ $\Sigma^{*+} \rightarrow\gamma\Sigma^+ $ & 0.210 & 0.233 & 0.303 \\ $\Xi^{*-}
\rightarrow\gamma\Xi^-    $ & 0.011 & 0.009 & 0.012 \\ $\Xi^{*0}    \rightarrow\gamma\Xi^0    $ & 0.313 & 0.300 &
0.415 \\
\end{tabular}
\caption{{\it Total decay widths normalized to that of the $\Delta\to\gamma N$ transition in various models. The
data for the quark model {\rm{(QM)}} and lattice calculation {\rm{(LATT)}} are taken from Refs.[6,7].}}
\end{center}
\end{table}

\section{Non-leptonic weak decays of octet hyperons}

To describe the non-leptonic hyperon decays within the present
approach we have to
introduce a weak effective lagrangian. We use the form\cite{DGH84}
\begin{equation}
\label{lag} {\cal L}_w = g \ Tr[ \lambda_6 \ \partial_\mu U
\partial^\mu U^\dagger ] +
      g' \ Tr[ \lambda_6 \ \partial_\mu U \partial^\mu U^\dagger
        \partial_\nu U \partial^\nu U^\dagger] +
     g'' \ Tr[ \lambda_6 \ \partial_\mu U \partial_\nu U^\dagger
     \partial^\mu U \partial^\nu U^\dagger]
\end{equation}
It should be noticed that this is not the most general lagrangian
one can write down up to fourth order in chiral counting. This general
form has been given in e.g. Ref.\cite{KMW90}. As well known, to second
order there is an extra $\Delta I=3/2$ term. Such $\Delta I=1/2$ rule violating
contributions, which also appear in next-to-leading order, are known to be small.
Consequently, they are neglected in what follows. Still, for
the decays we are interested in (no e.m. fields) there are 15
independent terms of fourth order. However, to fit the known data on
$K \rightarrow \pi\pi$ and $K \rightarrow \pi\pi\pi$, and assuming
complete octet dominance, it is enough to
consider only the three terms included in Eq.(\ref{lag}). Since in the meson sector
no other empirical information is available to determine the
low energy constants of a more general effective weak lagrangian we will just
concentrate on the form given in Eq.(\ref{lag}) and neglect
the remaining possible terms.

In our calculations we will consider the set of values\cite{DGH84}
\begin{equation}
g = 3.6 \times 10^{-8} \ m_\pi^2 \qquad ; \qquad g'/g = 0.15 \ fm^2 \qquad ;
\qquad g''/g = -0.067 \ fm^2
\label{values}
\end{equation}
for the constants $g$, $g'$ and $g''$. Slightly different values
are obtained from the fit done in Ref.\cite{KMW91}. However, since
both sets of parameters lead essentially to the same predictions
for the hyperon decay amplitudes we will report here only those
obtained using the values (\ref{values}).

To calculate the hyperon decays in the context of the collective
coordinates $SU(3)$ Skyrme model we should include the soft meson
fluctuations on top of the soliton background. This is achieved
using
\begin{equation}
U = \sqrt{U_M} \ U_0({\bf r}, t) \ \sqrt{U_M} \label{fullu}
\label{pp}
\end{equation}
where $U_M = 1 + i\ \vec \tau \cdot \vec \pi / f_\pi +...$.
Inserting Eq.(\ref{pp}) into ${\cal L}_w$, expanding the resulting
expression to linear order in the pion field and taking the
appropriate matrix elements one can finally obtain the corresponding
$S$-wave and $P$-wave decay amplitudes. This is worked out in
detail in the following two sections.

\subsection{S-waves amplitudes}

The $S$-wave amplitude $A$ for the process $B \rightarrow B'
\pi$ is obtained by
inserting Eqs.(\ref{ansatz}) and  (\ref{fullu}) in ${\cal L}_w$ and
taking matrix elements of the terms linear in $\pi$.
For the decay amplitudes with $\pi_0$ emission one gets
\begin{equation}
A(B \rightarrow B' \pi_0) = I_0  \ <\Psi_{B'}| D_{78} | \Psi_B >
\label{amp}
\end{equation}
where $I_0$ is a radial integral which depends on the soliton
profile and the constants $g$, $g'$ and $g''$. Its explicit
expression has been given in Ref.\cite{DGL85}.
In Eq.(\ref{amp}), $D_{ab}$ represents the $SU(3)$ rotation matrix
$D_{ab} = 1/2 \ Tr\left[ \lambda_a A \lambda_b A^\dagger \right]$.
Its matrix elements between the collective wavefunctions
corresponding to the different baryon states can be expressed as
linear combinations of $SU(3)$ Clebsch-Gordan coefficients and,
therefore, they can be easily calculated. Similar expressions are
obtained for the decays in which charged pions are emitted.

The results are summarized in the Table 2. Also listed in that table are the results of the quark
model\cite{DGH86}. In general our values are somewhat below those of the QM. However, in comparing with the
empirical results we see

\begin{table}
\begin{center}

\begin{tabular}[h]{cccc}
Transition                       &  This model  &  QM   &  Emp  \\ \hline $\Lambda^0 \rightarrow n \pi_0$  &
$-1.63$        & $-1.5$  & $-2.37$  \\ $\Sigma^+  \rightarrow p \pi_0$  & $-2.48$        & $-3.8$  & $-3.27$  \\
$\Sigma^-  \rightarrow n \pi_-$  &  3.50          &  4.2    &  4.27 \\ $\Xi^0     \rightarrow \Lambda \pi_0$  &
2.37          &  3.1    &  3.43 \\ $\Sigma^+  \rightarrow n \pi_+$  &    0           &   0     &  0.13 \\
\end{tabular}
\caption{{\it Calculated $S$-wave non-leptonic hyperon decay amplitudes as compared with the empirical values. For
comparison we also list the results of the quark model {\rm{(QM)}} taken from Ref.[8]. All values should be
multiplied by $10^{-7}$.}} \label{tasw}
\end{center}
\end{table}

\noindent that they are of similar quality. As in the case of the radiative decays it is interesting to consider
the ratio of amplitudes taken with respect to one of them. For those ratios the agreement between the calculated
and empirical values is indeed quite good. The vanishing value of $A(\Sigma^+ \rightarrow n \pi_+)$ is a direct
consequence of the octet nature of the ${\cal L}_w$ we are using. As already mentioned there are, in general, extra
terms that transform as 27-plets under flavor transformations. Those terms will be responsible for the small
violation of the $\Delta I = 1/2$ rule observed in the empirical values.

\subsection{P-wave amplitudes}

We turn now to evaluation of the $P$-wave amplitudes $B$. In the non-relativistic approximation they are defined by
\begin{equation}
{B\over{2 \bar M}} \ \chi^\dagger(\lambda') \ \vec \sigma \cdot \vec q \ \chi(\lambda)
\end{equation}
where $\vec q$ is the pion momentum, $\vec \sigma$ the spin operator and $\bar M$
is the average baryon mass.
In the present model $B$ receives two types of contributions\cite{DGL85}. One
is a contact contribution (see Fig.1a). The other type of contribution is given by the three
pole diagrams shown in Figs.1b-1d.

\begin{figure}[h]
\centerline{\psfig{figure=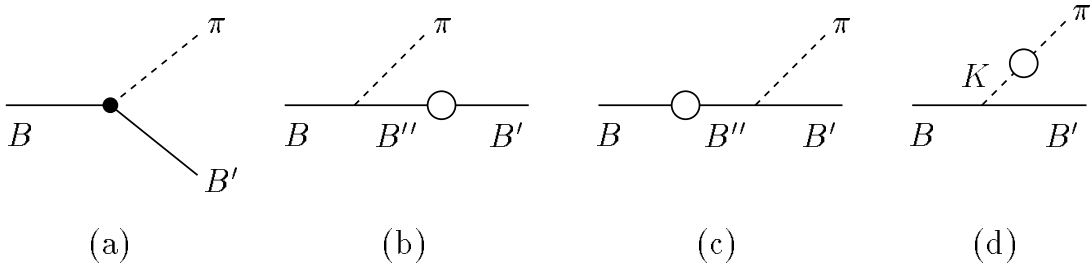,height=6.0cm}}
\caption[]{
{\it Diagrams contributing to the $P$-waves decays. Diagram (a)
represents the contact contribution. Diagram (b)-(c) are the
baryon pole contributions. Diagram (d) is the kaon pole
contribution.}} \label{f1}
\end{figure}

The contact contribution to the $P$-wave amplitude is obtained by
replacing Eqs.(\ref{ansatz}, \ref{fullu}) in ${\cal L}_w$ and
taking matrix elements of the terms proportional to $\partial_\mu \pi$.
For example, for the amplitudes with $\pi_0$ emission one finds
\begin{equation}
B_{contact}(B \rightarrow B' \pi_0) = \frac{2 ( M_B + M_{B'} )}{3 f_\pi}
<\Psi_{B'}|  \left(  I_1 D_{63} + I_2  D_{68} D_{33} + I_3 D_{63} D_{38}
\right) | \Psi_B >
\label{bcont}
\end{equation}
where $M_B$ and $M_{B'}$ are the masses of the initial and final
baryons, respectively. $I_1$, $I_2$ and $I_3$ are radial integrals of
rather involved functions of the soliton profile $F(r)$ which also depend on
the constants $g$, $g'$ and $g''$. Their explicit expressions can be found in
Ref.\cite{GGS99}\footnote{We believe that some of the expressions given
in the erratum of Ref.\cite{DGL85} contain errors
and/or misprints.}.

On the other hand, the total pole term contribution to those processes is
\begin{equation}
B_{pole}(B \rightarrow B' \pi_0) =
{M_B + M_{B'}\over{\sqrt{2} f_\pi}} \sum_{B''}
\left[ g_A^{B'B''}  \frac{A(B \rightarrow B'' \pi_0)}{M_B - M_{B''}} +
g_A^{B B'}  \frac{A(B'' \rightarrow B' \pi_0)}{M_{B'} - M_{B''}} \right]
\label{bpole}
\end{equation}
In deriving Eq.(\ref{bpole}) we have dropped a negligible small $K$ pole term contribution
(Fig.1d) and made use
of the generalized Goldberger-Treiman relations to relate the strong coupling constants
to the axial
charge. In the present model the axial charge operator has been
given in Ref.\cite{PSW91}.

The numerical results for the $B$ amplitudes in which a neutral pion is emitted are given in
Table 3. We observe that our predictions fail to reproduce the empirical decay amplitudes.
The same happens for the charged pion amplitudes.
In this sense, the situation is similar to that of other
models such as quark models or heavy baryon chiral perturbation theory.
For comparison some typical results of this latter approach are also listed in Table 3.
They correspond to a lowest order chiral calculation. Next-to-leading order
calculations do not improve the situation\cite{Jen92}. It is
important to note that in our model the pole contributions are, in
general, the result of a rather strong cancellation between the two
(large) baryon pole contributions (Figs.1b-1c). Consequently, they appear to be very
sensitive to the details of the model. In fact, something similar
happens in the alternative baryon model calculations. As for the
contact terms we observe that they are small and, therefore, do not
contribute very much to explain the observed $P$-wave amplitudes.

\section{Conclusions}

In this contribution we have presented some results for the radiative decay widths of the decuplet hyperons and the
non-leptonic decay amplitudes of the octet hyperons in the context of the collective coordinate approach to the
$SU(3)$ Skyrme model. We have found that, when normalized to that of the $\Delta N$ transition, the predictions for
the radiative decay widths are qualitatively similar to those of alternative quark based models. On the other hand,
the calculated $\Delta N$ transition amplitude turns to be roughly $30 \%$ smaller than the empirical value. In the
context of the $SU(2)$ Skyrme model it has been shown\cite{MW97}, however, that the inclusion of next-to-leading
order quantum corrections solves this problem. In the $SU(3)$ model such corrections have been already included in
the evaluation of the baryon masses\cite{SW98} and corresponding results for the radiative decays widths will be
available soon. In any case, it is important to stress that they are not expected to affect the ratios of decay
amplitudes presented here. The situation concerning the non-leptonic weak hyperon decays is, unfortunately, not
very satisfactory. Although we find that the model provides a rather good description of the empirical $S$-wave
decay amplitudes, it fails to reproduce the observed $P$-wave ones. In particular, the contact contributions to the
$P$-waves are not sufficient to explain the discrepancies between the pole contributions and the empirical values.
Thus, in contrast to previous expectations\cite{DGH86,DGL85}, within the present approximations the Skyrme model
does not seem to provide a solution to the long-standing '$S$-wave/$P$-wave puzzle' which affects the available
model calculations.

\section*{Acknowledgements}

The material presented here is based on work done in collaboration with A. Garcia, D. G\'omez Dumm,
T. Haberichter, H. Reinhardt and H. Weigel. Support provided by the grant PICT 03-00000-00133 from
ANPCYT, Argentina is acknowledged. The author is fellow of the CONICET, Argentina.
He would like to thank the members of the Organizing Committee for their warm hospitality
during his stay in Erice.

\vspace*{2cm}

\begin{table}[b]
\begin{center}

\begin{tabular}[h]{crrrr}
Transition   &\multicolumn{2}{c}{This model}         &   Chiral Fit   &  Emp
                \\ \cline{2-3}
                                 &  pole   & contact   &         &          \\ \hline
$\Lambda^0 \rightarrow p \pi_0$        & $-5.15$   &  $-0.38$    &   $-16 \qquad $  &  $-15.8$ \\ $\Sigma^+
\rightarrow p \pi_0$        &  2.73     &  2.23       &   $10  \qquad $  &    26.6  \\ $\Xi^0     \rightarrow
\Lambda \pi_0$  &  1.98     &  $-0.27$       &   $ 3.3 \qquad $ &  $-12.3$ \\
\end{tabular}
\caption{{\it Calculated $P$-wave non-leptonic hyperon decay amplitudes as compared with the empirical values. For
comparison the results of a lowest order chiral fit taken from Ref.[23] are also listed. All values should be
multiplied by $10^{-7}$. }} \label{tapw}
\end{center}
\end{table}

\end{document}